# Chua Mem-Components for Adaptive RF Metamaterials

J. Georgiou, K. M. Kossifos and M. A. Antoniades
Dept. of Electrical and Computer Engineering
University of Cyprus
Nicosia, Cyprus
Email: julio@ucy.ac.cy

A.H. Jaafar and N. T. Kemp
School of Mathematics and Physical Sciences
University of Hull
Cottingham Rd, HU6 7RX, Hull, United Kingdom
Email: n.kemp@hull.ac.uk

*Abstract*—Chua's mem-components are ideal for creating adaptive metasurfaces for manipulating EM waves given that they hold their state without external biases. In this paper, we propose a generic adaptive reactive element that is in fact a memcapacitor/meminductor. This element makes use of a polymer that demonstrates reversible trans-cis photochemical isomerization, thus making it possible to change the distance between two conductive plates by up to 25%. Furthermore, a design methodology for utilizing these devices is presented.

## I. INTRODUCTION

Transistor scaling has been the most important driving force behind improved performance of electronics to-date, however with the end of Moore's law approaching [1], a more-than-Moore approach is urgently required i.e. there is a need to start looking for and utilizing new devices that in combination with current technology can achieve greater computational power. In 1971, Leon Chua observed from circuit theory, through examining the relationships between the primitives: current *i*, charge *q*, voltage *v* and flux *φ*, that there was a passive component missing needed to link charge *q* with flux *φ* [2]. It led to the formalization of the *memory varistor/resistor* or "memristor" in short, where:

$$M(q) = \frac{d\phi(q)}{dq} \quad (1)$$

This work went largely unnoticed by industry until 2008, when HP researchers [3][4] announced the first commercially intent devices. Since then the memristor has progressed rapidly and today it has found its way into commercial products, demonstrating high-performance and incredibly dense non-volatile, memristor-based memories [5]. More recently an optical switching memristor has been demonstrated [6]. Furthermore, the "Chua mem-device" concept was recently extended [7] to define the memcapacitor and the meminductor, where, the capacitance or inductance values respectively depend on inputs of the past.

Although at an early stage in their development, solid-state memcapacitors have been suggested [8] in the past. These particular memcapacitors are based on multilayer capacitive structures, which allow tunneling between layers for charge transport. Furthermore, a solid state meminductor has also been demonstrated [9]. Thus with the increasing availability of mem-components, one has access to adaptive, non-volatile, "resistors, capacitors and inductors", that offer the possibility of compact arrays of adaptive passive elements, when organized in a cross-bar structure [10]. An application area where Chua's mem-components have the potential to be a disruptive technology is that of metamaterials and metasurfaces, since these require relatively dense arrays of components, especially as the frequencies of operation increase. Metamaterials (MTMs) are periodic structures that rely on LCR resonators, with the extent of inductance, capacitance and resistance required, depending on the function desired from the material, e.g. if perfect anomalous reflection is desired then the R component should be minimized and the focus should be on attaining the correct values for the reactive components L and C, for a particular frequency and angle of incidence. On the other hand, if a perfect absorber is required then a tunable R is extremely important in being able to maximally absorb the incoming radiation. Currently, MTMs and metasurfaces are created either with printed or lumped elements, with lumped elements [11] being recently used to design a super-lens. This was achieved by designing the MTM to possess electric permittivity and magnetic permeability equal to −1. Split ring resonators (SRRs) and closed ring resonators (CRR) are among many types of printed forms of electromagnetic resonators that are also widely used in MTMs. SRRs were used in [12] to cloak a metallic object from an electromagnetic wave. This was done by creating a progressive change in refractive index (RI), around the object, such that the electromagnetic wave propagated around the cloaked object and didn't interact with it. Anomalous reflection [13] and absorption [14] was also obtained by MTMs while maintaining their planar geometry. The issue with the MTM-based metasurfaces is that the structures are highly dependent on the incident frequency, as well as the angle on incidence. Thus, if adaptive metasurfaces are to be created, it is necessary to have compact, adaptive R, L and C elements, which do not require a continuous bias voltage to maintain their properties, therefore enabling a single controller to modify them. Chua mem-components, embedded behind or even within the metasurfaces, as a cross bar structure are a promising way forward, since a controller on the periphery could manage the adaptation of the elements embedded within a metasurface, without the requirement of each meta-atom having its own controller.

In this paper, we propose a general reactive mem-component, that is capable of giving both memcapacitive and meminductive behavior. It is suitable to be embedded behind a metasurface structure, with the internal state of the reactive component being controlled through an optical signal, derived from charge or flux flowing through the device.

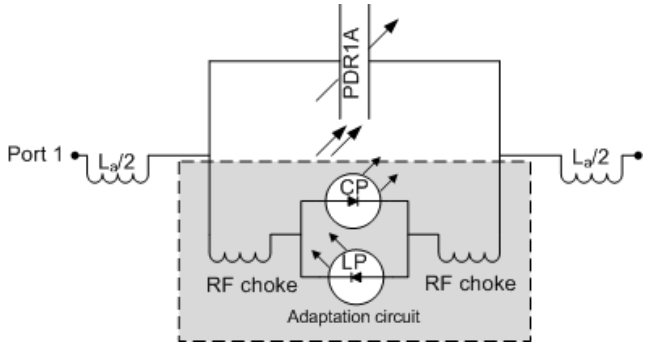

**Figure 1-Proposed Reactive Memcomponent-** RF chokes are included so as to block the RF signal from traversing through the tuning section, thus the bottom half of the circuit appears like an open circuit at the frequencies of interest.

## II. PROPOSED REACTIVE MEM-COMPONENT

### A. Overview of Reactive Mem-Component

Figure 1 shows an overview of the proposed reactive mem-component. It consists of a capacitor whereby the dielectric separating the plates of the capacitor is constructed using a layer of *poly disperse red 1 acrylate* (PDR1A). How this layer changes the capacitance will be explained in the next section. Furthermore, the component consists of two diode-based light sources, where one produces linearly polarized light and the other produces circularly polarized light, designed to interact with the PDR1A layer, which is located directly above. Interaction with the PDR1A layer is enabled by a transparent, conductive indium tin oxide layer (ITO), which forms one of the plates of the capacitor. In order to prevent the RF signal of interest interacting with the diodes, two RF chokes isolate the diodes from the signal path. Finally, two predominantly inductive tracks connect the variable capacitor to the ports of the element.

### B. The Active Structure and PDR1A Properties

The key to the adaptive capacitor is the PDR1A dielectric. Its mechanics have been described in [15][16]. This material expands when exposed to circularly polarized (CP) light and contracts when exposed to linearly polarized (LP) light. The azobenzene units that comprise it undergo reversible trans-cis photochemical isomerization upon illumination. In its bulk or thin film form, optical irradiation induces a large photomechanical response by progressive alignment of chromophores due to repeated trans-cis-trans cycling. The volume of PDR1A can expand or contract by 25% depending on the current state and which kind of polarization of light is incident. This expansion/contraction will cause a capacitance variation of about 20%.

After the substrate expanded and the CP light is removed the substrate can return to its initial state. However, when exposed to LP light, the inherent thermal relaxation time of the material is eight times shorter, than without it. This hysteresis is what causes the memory effect of the device.

## III. DESIGNING WITH A GENERIC REACTIVE MEM-COMPONENT

Figure 2 illustrates a particular design of the capacitive structure, which is used in the analysis to follow. In this design, two substrates are used, a high frequency substrate ($Sub_2$) and PDR1A on top ($Sub_1$). The structure can be viewed as a parallel plate capacitor and two transmission lines arising from the tracks leading to it, with width $W_a$, over the ground plane. The top plate of the capacitor is made of a transparent conductor indium tin oxide (ITO) so that the dielectric ($Sub_1$) can be illuminated underneath when the device is exposed to light. We have made the simplifying assumption that the frequency of interest is blocked by the RF chokes leading to the adaptation light sources. The structure $L_2$ is the device length and, $L_1$ is the length where the top and bottom plate overlap.

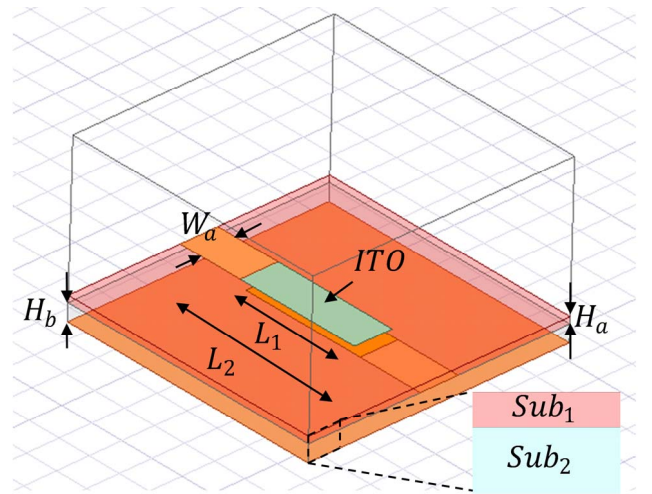

**Figure 2 -** Mem-component. $Sub_1$ is PDR1A and $Sub_2$ is RO4003. Dimensions: $H_a = 25\ \mu m, H_b = 0.254\ mm$

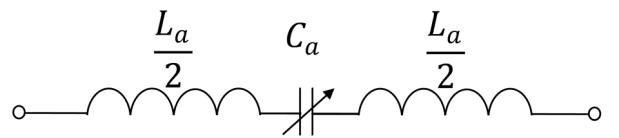

**Figure 3 -** Mem-component RF signal path circuit with inductor elements.

This geometry can be modeled as a series variable capacitance and two inductors. It is widely known that for series elements the total impedance of the components is the summation of all the element impedances. In Figure 3, we can see the equivalent circuit of the structure of Figure 2. The total impedance for this simple circuit is purely reactive and is equal to:

$$X_{MEM}(q) = \omega L_a - \frac{1}{\omega C_a(q)} \quad (2)$$

where:

$$C_a(q) \approx \varepsilon_0 \varepsilon_r \frac{W_a L_1}{H_a\ a(q)} \quad (3)$$

$a$ is the state variable and since the PDR1A substrate can expand 25% the coefficient can take values from 1 to 1.25.

$$1 \leq a \leq 1.25 \quad (4)$$

The value of $a$ depends on the duration of illumination and polarization of the PDR1A substrate and therefore the charge and polarity that has passed from diodes shown in Figure 1. From the above one can say that:

$$a(q) = 1 + max\left\{min\left\{k_a \int_{t_0}^{t} q(\tau)d\tau \Big| 0.25\right\} \Big| 0\right\} \quad (5)$$

The term $k_a$ is an expansion parameter, relating the expansion to the charge. This equivalent circuit shown in Figure 3 assumes that the physical dimensions of the component are negligible, much smaller than the wavelength. For higher frequencies where metamaterials usually operate the equivalent circuit in Figure 4 could be used. For the equivalent circuit in Figure 4 the reactance is equal to:

$$X_{MEM}(q) = \frac{(2\omega Z_0 C_a a(q) - \cot{\theta/2})\sin\theta}{2\omega C_a a(q)} \quad (6)$$

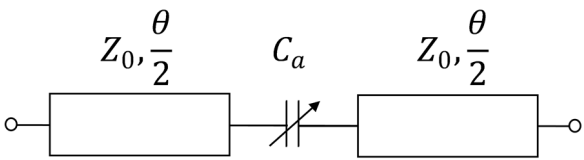

**Figure 4 - Mem-component circuit with transmission line elements.**

where $Z_0$ is the transmission line impedance and $\theta$ is the electric length of the line. In examining equation (2), one can conclude that, depending on the frequency, the capacitance value ($C_a$) and inductance values ($L_a$), the total reactance could be positive (inductive), zero (at resonant frequency) or even negative (capacitive). The same holds for (6) depending on $C_a$, $Z_0$ and $\theta$, the total reactance could be positive negative or zero.

The structure shown in Figure 2 was simulated using the ANSYS HFSS electromagnetic simulator, and the resulting reactance is shown in Figure 5. For that particular geometry $C_a = 1.3\ pF$ and $L_a = 0.644\ nH$, $Z_0 = 50\ \Omega$, and $\theta = 25.1°$. It can be seen that the reactance is capacitive below 5.5 GHz, zero at 5.5 GHz, and inductive above 5.5 GHz. It should be noted that it is best to operate the device away from its self-resonant frequency in order to avoid large fluctuations in the value of the reactance.

The design flow for establishing desired properties, based on the structure in Figure 2 requires that one first chooses $C_a$ and $Z_0$ and then determines $W_a$ in order to obtain the input impedance. $C_a$ is dependent on the Substrate 1 ($Sub_1$) thickness ($H_a = 25\ \mu m$) and on the area of the capacitor ($W_a \times L_1$). Thus, solving (6) for the reactance needed, in this case $X_{MEM} = 0$, one can obtain $\theta$. The solution is periodic, therefore we hold the smaller realizable solution. Then $\theta$ is converted to a physical length $L_2$.

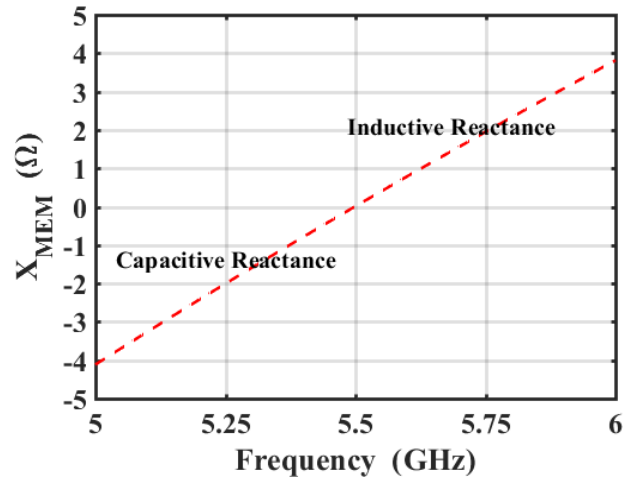

**Figure 5 - Mem-component reactance. $X_{MEM} = 0$ at 5.5 GHz. Dimensions (mm): $W_a = 0.58$, $L_2 = 2.17$, $L_1 = 1.8$.**

In Figure 6 the capacitance of the proposed mem-component, operating in the capacitive region, is shown. The component in this case was designed by keeping $C_a$ fixed at $C_a = 1.3\ pF$ and reducing the overall length of the device. In order to reduce the physical length of the device the width is increased to ($W_a = 0.81\ mm$). This had the effect of reducing the transmission line impedance ($Z_0$) to 40 Ω. With the area of the device kept constant for the given width, ($W_a$), $\theta$ can be calculated to obtain the desired capacitance. In the same figure the capacitance is plotted for multiple expansion states of PDR1A to demonstrate the effect of the capacitance change over the whole mem-component.

In a similar manner to that shown in Figure 6 the design can be adapted, just by varying the drawn geometries, to operate the adaptive device in the inductive state, i.e. to have a Meminductor. An inductive component response can be seen in Figure 7. In this case, again the memcapacitance component is kept constant but the length of the device is increased.

Thus it is shown that by changing the width ($W_a$) of the capacitive component of the device and the length $L_2$-$L_1$ one can convert the memcapacitor into a meminductor. Through the polarity and duration of the DC signal that controls the LP and CP light sources, one may make fine adjustments on the fly, enabling not only a high degree of control but as well a mechanism, when combined with a suitable form of feedback, a device that can reconfigure and adapt autonomously to desired changes or environmental responses.

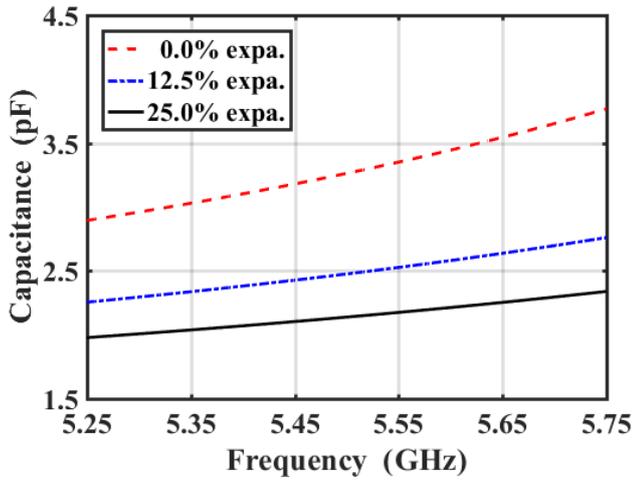

Figure 6 - Mem-component acting as a memcapacitor. Dimensions (mm): $W_a = 0.81, L_2 = 1.49, L_1 = 1.29$.

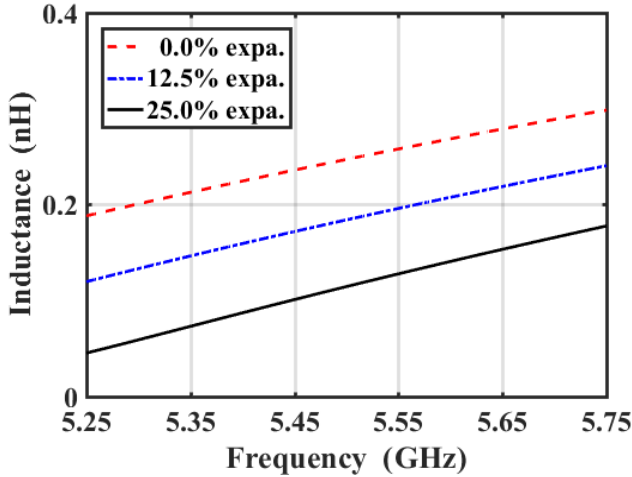

Figure 7 - Mem-component acting as a meminductor. Dimensions (mm): $W_a = 0.44, L_2 = 2.57, L_1 = 2.37$.

## IV. ADAPTATION

A DC current flowing from port 1 to port 2 in Figure 1, has the effect of lighting up the CP light source, hence increasing the distance between the capacitive plates and decreasing the overall inductance of the composite device, by decreasing the capacitive component. Similarly, a negative DC current flowing from port 1 to port 2 has the effect of lighting up the LP light source, hence decreasing the distance between the plates and thus decreasing the overall inductance.

$$\frac{dL}{dt} = k_1 \times I_{diode} = k_1 \times \frac{dq}{dt} \quad (7)$$

where $k_1$ is a constant relating the change of L to the current through the diode.

Similarly, in an overall capacitive structure the following holds:

$$\frac{dC}{dt} = -k_2 \times I_{diode} = -k_2 \times \frac{dq}{dt} \quad (8)$$

where $k_2$ is a constant relating the change of C to the current through the diode.

A very slow AC signal will increase and decrease the reactance of the reactive mem-component, however as the frequency is increased beyond a particular point, the capacitance/inductance will remain in its current state, thus giving a linear relationship between dq and dv, and dφ and di respectively.

## V. CONCLUSION

In this paper a reactive mem-component has been presented that is suitable for incorporation in metamaterials and metasurfaces. It essentially consists of a memcapacitor that is fabricated using PDR1A as a dielectric, which is tuned by a DC (or very low frequency) voltage or current driven through a linearly polarized or circularly polarized light source. If a meminductor is required, a series inductance can be created by changing the aspect ratio of the capacitor to make it narrower in addition to increasing the length of the series connected track length. The meminductance is tuneable by adjusting the incorporated memcapacitance.

## ACKNOWLEDGMENT

This work is partially funded from the European Union's Horizon 2020 Programme FETOPEN-2016-2017, through the VISORSURF project, under grant agreement no. 736876. The authors would like to acknowledge support by the COST Action IC1401.